\documentclass[12pt,a4paper]{revtex4}
\usepackage{multirow}
\usepackage{amssymb}
\usepackage{amsmath,graphicx}
\usepackage{array}
\usepackage{braket}
\usepackage{bm}
\usepackage{enumerate}
\usepackage{epstopdf}
\usepackage{amsmath}
\usepackage{dcolumn,caption,booktabs}
\usepackage{slashed}
\usepackage{amsfonts}
\usepackage{adjustbox}
\usepackage{tcolorbox}
\usepackage{float}
\usepackage{collectbox}
\usepackage{listings}
\usepackage[unicode]{hyperref}
\usepackage[mathscr]{eucal}
\usepackage{graphics}
\usepackage{subcaption}
\newcommand{\be}{\begin{equation}}
	\newcommand{\ee}{\end{equation}}
\newcommand{\bea}{\begin{eqnarray}}
	\newcommand{\eea}{\end{eqnarray}}
\newcommand{\ben}{\begin{enumerate}}
	\newcommand{\een}{\end{enumerate}}
\newcommand{\bde}{\begin{widetext}}
	\newcommand{\ede}{\end{widetext}}

\newcommand{\al}{\alpha}
\newcommand{\la}{\lambda}

\newcommand{\om}{\omega}

\newcommand{\fr}{\frac}

\newcommand{\bc}{\begin{center}}
	\newcommand{\ec}{\end{center}}

\newcommand{\ep}{\epsilon}

\newcommand{\ka}{\kappa}
\newcommand{\La}{\Lambda}

\newcommand{\Om}{\Omega}

\newcommand{\betah}{\frac{\beta}{H*}}

\begin{document}
\newcommand{\AdrHEPC}{$^a$Department of Theoretical Physics, University of Science, Ho Chi Minh City 700000, Vietnam\\ $^b$Vietnam National University, Ho Chi Minh City 700000, Vietnam}	
	\title{Sphaleron and gravitational wave with the Higgs-Dilaton potential in the Standard Model Two-Time Physics}
	
	\author{Vo Quoc Phong$^{a,b}$}
	\email{vqphong@hcmus.edu.vn}
	\affiliation{\AdrHEPC}
	\author{Quach Ai Mi$^{a,b}$}
	\email{aimiwu14498@gmail.com}
	\affiliation{\AdrHEPC}
	\author{Nguyen Xuan Vinh$^{a,b}$}
	\email{vinhnguyen.mxt@gmail.com}
	\affiliation{\AdrHEPC}
\begin{abstract}
By introducing a Higgs-Dilaton potential, the 2T model has a trigger for a first order electroweak phase transition, namely for the mass of Dilaton between $300 $ GeV and $550$ GeV. We have also compared the transition strengths in the case with and without daisy loops, the difference being always less than $0.2$. The effective Higgs potential has given a sphaleron energy less than $8.4$ TeV. The timescale of phase transition $(\beta/H^*)$ is larger than $25$ and less than $34$ in all cases that are sufficient to trigger the first order electroweak phase transition. Gravitational wave energy density caused by this transition, may be detected by future detectors, could indirectly confirm Dilaton.

\end{abstract}
\pacs{11.15.Ex, 12.60.Fr, 98.80.Cq}
\maketitle
Keywords:  Spontaneous breaking of gauge symmetries,
Extensions of electroweak Higgs sector, Particle-theory models (Early Universe)

\section{INTRODUCTION}\label{secInt}

Multidimensional theories have become an important phenomenological research framework. However, from 2008, besides the string theory, the multi-dimensional model that fully combines with particle physics is currently only the 2T theory \cite{bars1999,bars2000a,bars2001,survey,bars2006,kuo2006,33}. The one not only provides us with a new view of time but also suggests new explanations for difficult problems such as the matter-antimatter asymmetry and the strong-CP problem. The cause of them may be from an exotic particle in the 2T model, Dilaton.

The 2T model is a form of space-time extension but at the same time proposes a new particle, Dilaton which is originally a dark matter candidate. However, the role of Dilaton could be more numerous. It as previously investigated is a source of strong CP violation and is likely also responsible for the matter-antimatter asymmetry.

The first order electroweak phase transition (EWPT) problem was solved in the 2T elementary particle model. According to current research, the triggers can be Dark matter (DM) or new heavy bosons  Refs.~\cite{5percent,sakharov,BSM1,BSM2,BSM3,majorana,thdm1,thdm2,ESMCO,phonglongvan1,phonglongvan2,phonglongvan3,phonglongvan4,dssm,munusm,lr,ppf1,ppf2,ppf3,r331a,r331b,r331c,r331d,r331e,Buras:2012dp,1101.4665a,1101.4665b}. As in the studies of other authors, the cause of first order phase transition is due to the dominant activation of Dilaton which is stored in an external potential of Dilaton.

Dilaton is a common name. The concept of Dilaton also appears in many other theories such as the composite Higgs models. However, Dilaton in the 2T model is a particle associated with the concept of two time dimensions and the symmetry $SP(2,R)$.

At present, the evidence for extra dimensions is very weak. Instead of looking for it, the 2T model needs to be enabled and look for the evidence of Dilaton. This also indirectly proves the existence of extra dimensions. The electroweak phase transition may be the clue for this search. Because the problem can give us the possible mass domain of Dilaton, sketching a form of Dilaton potential. In addition, it makes clearer the mass generation scenario for particles.

In Ref.\cite{pn}, we proposed a form of Dilaton potential with the non-zero mass Dilaton which is sufficient to trigger a first order electroweak phase transition. However, in Ref.\cite{pn} with a one-loop effective potential without daisy loops, the necessary mass domain for Dilaton is given but not completely. In this paper, an effective potential with daisy loops is used to recalculate more accurately the phase transition strength ($S$) and other quantities in the Baryogenesis scenario. Specifically the sphaleron energy and the gravitational wave energy (GW) density are calculated.

This article is organized as follows. In section \ref{sec2}, the types of Dilaton potentials are discussed. Then the strength of EWPT is calculated in the case of a one-loop effective potential with Daisy loops. The comparison of phase transition strength in the case with and without daisy loops. In section \ref{sec3}, by using static fields and spherically symmetric bubble nucleations, sphaleron energy and gravitational waves are solved numerically with a wide range of Dilaton mass. We make a general assessment of sphaleron energies, comparing the gravitational wave energy density with current data. Finally, we summarize and make outlooks in section \ref{sec4}, such as evaluating the compatibility of sphaleron with other conditions and giving a way to search for an first order electroweak phase transition.

\section{Controling EWPT by the Higgs-Dilaton Potential}\label{sec2}

The basic concepts as well as the 2T elementary particle model of 2T theory were built by I. Bars \cite{bars2006} when considering multidimensional concepts. In our opinion, it is like an extension of the concept of time. This model has successes such as solving the problem of strong CP violation, providing DM candidates \cite{bars2001}.

The SM in the 2T model includes the Higgs scalar fields $H$ and Dilaton $\Phi$, the left/right chiral fermion fields $\Psi^L, \Psi^R$ (which are quarks and leptons), and the gauge bosons $A_M$ that accompany the gauge symmetry of this model \cite{bars2006}.

With the $SP(2,R)$ symmetry, the Higgs-Dilaton potential \cite{kuo2006} has the following form,
\begin{equation}
V(\Phi,H) = \la\left(H^\dagger H - \al ^2 \Phi^2\right)^2 + V(\Phi),\label{47v}
\end{equation}
in which $\la, \al $ are dimensionless couplings. The Dilaton $\Phi$ and the Higgs doublet $H$ which are the $SO(4,2)$ scalar fields \cite{bars2006},

\begin{equation}
	\Phi=\left( 1,1\right)
	_{0},\;\;H=\left(
	\genfrac{}{}{0pt}{}{H^{+}}{H^{0}}	
	\right)  _{\frac{1}{2}}=\left(  1,2\right)  _{\frac{1}{2}}.\nonumber
\end{equation}

Initially $V(\Phi)$ had no specific form.  But it is the key to examining EWPT and related quantities \cite{1bars}. The minimization process Eq. \eqref{47v} leads to the Higgs and Dilaton fields having non-zero VEV and being proportional to each other \cite{pn, bars2006}.

After reducing to 1T \cite{bars2006,phong2022}, we get the 4D Higgs-Dilaton potential as follows:
\begin{align}\label{47vb}
V(h,\phi)= \la\left[\chi^2 - \al ^2\phi^2\right]^2 + V(\phi), \chi(x) = \dfrac{1}{\sqrt{2}}(v + h(x)), \phi=\dfrac{1}{\al \sqrt{2}}(v + \al  d(x)),
\end{align}
in which $v=246$ GeV, $h, d$ scalar fields with zero VEVs. When $v(\phi)$ is involved, $h,d$ are not physical fields yet. Because $V(h,\phi)$ will contain mixed components between $h$ and $d$. Depending on the form of $V(\Phi)$, the diagonalization process will yield physical particles.

Because of the $SP(2,R)$ symmetry, in 6-dimensional spacetime, the potential of scalar field must have a form of fourth powers of field. Therefore, the Dilaton potential is rewritten in its general form as $f(\mathcal{S}/\Phi).\Phi^4$, from which we see that the potential must always be the even power of field. The Dilaton field in the 2T model with a potential of only fourth power, is like a degree of freedom or it has near zero mass but has a non-zero VEV. 

So this is something different from SM. However, the electroweak symmetry breaking is accompanied by the $SP(2,R)$ symmetry breaking. When this symmetry is broken, the Dilaton potential is no longer only of fourth power. This can be explained mathematically by choosing the function $f(\mathcal{S}/\Phi)$.

The form of $V(\Phi)$ is still under consideration in some cases, for example in connection with gravity. In Table \ref{phanloai} the choices of $V(\Phi)$ are summarized.

\begin{table}[!ht]
	\centering	
	\begin{tabular}{c|c|c|c|c}\hline\hline
		$f(\mathcal{S}/\Phi)$&$V(H,\Phi)$&$\mathcal{S}$& $V(\Phi)$ & Refs.\\ \hline
		\multicolumn{5}{c}{$V(H,\Phi)=f(\mathcal{S}/\Phi)\Phi^4$}\\
		\hline	
		$\frac{\lambda}{4}(\mathcal{S}^2/\Phi^2-\alpha^2)^2+\rho/4$&$\la\left(H^\dagger H - \al ^2 \Phi^2\right)^2+\rho \Phi^4$& $\sqrt{2H^\dagger H}$ & $\rho \Phi^4$& Refs.\cite{1bars}\\		
		$\frac{\lambda}{4}(\mathcal{S}^2/\Phi^2-\alpha^2)^2+\rho/4+\frac{\omega^2}{\kappa^2\Phi^2}$&$\la\left(H^\dagger H - \al ^2 \Phi^2\right)^2+\rho \Phi^4+\frac{\omega^2}{\kappa^2}\Phi^2$& $\sqrt{2H^\dagger H}$ &$\rho \Phi^4+\frac{\omega^2}{\kappa^2}\Phi^2$ & Refs.\cite{pn,1bars}\\
		\hline\hline
	\end{tabular}
	\caption{Cases of the Higgs-Dilaton potential}\label{phanloai}
\end{table}	

\subsection{The Dilaton potential}

In this study, an added Dilaton potential has the following form \cite{pn}:

\begin{equation}
V(\Phi) = \rho \Phi^4 - \dfrac{\om ^2}{\ka ^2}\Phi^2.\label{1p}
\end{equation}

$\omega$ is much smaller than $\kappa$. $\omega$ has the mass dimension, $\kappa$ is the scaling constant when reducing from the 2T to 1T spacetime, $\kappa$ is a mixture of the extra dimension and the second time dimension so it has the length dimension \cite{1bars,bars2006}. The second power component of field in Eq.\ref{1p} is a soft-breaking component of the $SP(2,R)$ symmetry. It implies the second power component of field, $\frac{\omega^2}{\kappa^2}\phi^2$ is very small compared to the fourth power component. As in Ref.\cite{pn}, $V(\Phi)\sim const.\Phi^4+const.\Phi^2$ has been assumed. In general, we can choose $\kappa$ to be arbitrarily large, so the above inequality can always be satisfied. After reducing from 2T to 1T, the potential will be in the form,

\begin{equation}
V(\phi) = \dfrac{\rho}{\ka ^4}\phi^4 -
	\dfrac{\om ^2}{\ka ^4}\phi^2.\label{thephi}
\end{equation}

Therefore we introduce only even-order components of the Dilaton field. This assumption could be justified if one takes into account the next leading order of 2T metric in the action \cite{bars2006} and $Z_2$ symmetry of Dilaton. $\kappa^4$ will be simplified by the 2T action when reduced to $1T$ \cite{bars2006}. Then by substituting the above Higgs and Dilaton fields into the Higgs-Dilaton potential, the potential is rewritten in terms of the variable $\phi_c$ which is a re-notation of VEV,
\begin{align} \label{50}
V(h,\phi) &= \dfrac{\lambda}{4}\left[(\phi_c + h)^2 - (\phi_c + \alpha d)^2\right]^2 - \dfrac{\omega^2} {2\alpha^2}(\phi_c + \alpha d)^2 + \dfrac{\rho}{4\alpha^4}(\phi_c + \alpha d)^4 \nonumber\\
&=\left\{\left(\dfrac{\rho}{4\alpha^4}\phi_c^4 - \dfrac{\omega^2}{2\alpha^2}\phi_c^2\right) + \left(\dfrac{\rho}{\alpha^3}\phi_c^3-\dfrac{\omega^2 }{\alpha}\phi_c\right)d\right.\nonumber\\
&\qquad\left.-2\lambda\alpha\phi^2_chd + \lambda\phi_c^2 h^2 +\left(\lambda\alpha^2\phi_c^2 - \dfrac{\omega^2}{2} + \dfrac{3\rho}{2\alpha^2}\phi_c^2 \right)d^2\right\} + \text{interaction terms},
\end{align}

Eq.\eqref{50} appears the term $hd$ so they are not physical particles yet. In the 2T spacetime, the $\omega^2/\kappa^2\Phi^2$ component will be so small that it can be neglected. However, when reducing the spacetime to 1T, the $\kappa^4$ parameter in Eq.\eqref{thephi} is lost, so the $\omega^2\phi^2$ component can have a significant contribution and lead to a Dilaton with non-zero mass.

The diagonalization process requires that the components proportional to the field $d$ must cancel. Therefore $\left(\frac{\rho}{\alpha^3}\phi_c^3-\frac{\omega^2 }{\alpha}\phi_c\right)=0$, which leads to $\omega^2=\frac{\rho}{\alpha^2}\phi^2_c$. This relation can not hold in the first order phase transition. It changes with $\phi_c$. However, the mathematical form is not likely to be preserved in the electroweak phase transition. This is similar to how some authors have modified the Yukawa interaction which leads to a first order electroweak phase transition (as shown in Ref.\cite{baldes}). But in the simplest of considerations, we assume that this form does not change. This is similar to considering that the mathematical form of the masses of particles may not change. But in the basic sense, this form can be considered unchanged. If we change this form, the interaction between the Higgs and the Dilaton will change very strongly, which requires the assumption of some other mechanism.\\

After diagonalization, we obtain physical particles which have masses at tree level as \cite{pn}
\begin{align}
m^2_{d'}=&\frac{2 \alpha^4 \lambda  \phi_c^2+2 \alpha^2 \lambda  \phi_c^2+3 \rho  \phi_c^2-\alpha^2 \omega^2}{2 \alpha^2}\nonumber\\
&-\frac{\sqrt{8 \alpha^2 \lambda  \phi_c^2 \left[\alpha^2 \omega^2-3 \rho  \phi_c^2\right]+\left[\phi_c^2 \left(2 \alpha^4 \lambda +2 \alpha^2 \lambda +3 \rho \right)-\alpha ^2 \omega ^2\right]^2}}{2 \alpha ^2},\label{gt1}\\
m^2_{h'}=&\frac{2 \alpha^4 \lambda  \phi_c^2+2 \alpha^2 \lambda  \phi_c^2-\alpha^2 \omega^2+3 \rho  \phi_c^2}{2 \alpha^2}\nonumber\\
&+\frac{\sqrt{8 \alpha^2 \lambda \phi_c^2 \left[\alpha ^2 \omega^2-3 \rho  \phi_c^2\right]+\left[\phi_c^2 \left(2 \alpha^4 \lambda +2 \alpha^2 \lambda +3 \rho \right)-\alpha^2 \omega^2\right]^2}}{2 \alpha ^2}.\label{gt2}
\end{align}

In Eq. \eqref{gt1}, as $\omega$ and $\rho$ go to zero, $m^2_{d'}=0$, which is in agreement with the results in Ref. \cite{bars2006}. The above formulas can be rewritten as follows
\begin{align}
m_{h'}^2(\phi_c) = A \phi_c^2; m_{d'}^2(\phi_c) &= B\phi_c^2,\label{higgs}
\end{align}
here $A,B$ are the parameteres. As $\phi_c$ approaches 0, $m_{d'}$ and $m_{h'}$ actually approach 0 in Eqs. \eqref{gt1} and \eqref{gt2}. According to the approximation formulas (Eqs. \eqref{higgs}), when $\phi_c \longrightarrow 0$, $m_{d'}\longrightarrow 0$ and $m_{h'}\longrightarrow 0$.

For convenience, $h',d'$ will be denoted again as $\mathcal{H},\mathcal{D}$ in the following sections. The contribution of extra dimensions does not only appear in one effective potential via Dilaton but also in the mass of Higgs and other fields as well.

Thus, Dilaton has a very interesting property, the manifestation of Dilaton in 1T spacetime is a massless particle. Or in other words, the manifestation of 6-dimensional spacetime (2T spacetime) is through the mass of Dilaton.

Furthermore, from the 2T sectors reduced to 1T, analyzed and diagonalized, the masses of top quark and gauge fields are obtained \cite{bars2001, bars2006}, respectively,
\begin{align}
	m_t^2(\phi_c) &= \dfrac{h_t^2}{2}\phi_c^2 = \dfrac{m_t^2}{v^2}\phi_c^2\,,\\
	m_W^2(\phi_c) &= \dfrac{g^2_1}{4}\phi_c^2  = \dfrac{m_W^2}{v^2}\phi_c^2\,,\\
	m_Z^2(\phi_c) &= \dfrac{g^2_1+g_2^2}{4}\phi_c^2 = \dfrac{m_Z^2}{v^2}\phi_c^2\,,\label{z}
\end{align}
where $m_i \equiv m_i(v)$ is the mass of field $i$ at the zero temperature and $v=246$ GeV, is equal to the VEV of Higgs in SM. As the temperature drops to zero, the VEV of the Higgs field increases to 246 GeV. Here $\phi_c$ is understood as VEV at temperature $T$. Therefore, the change from the symbol $v$ to $\phi_c$ in the previous steps is convenient for writing expressions at temperature $T$. From Eq.\eqref{higgs} to Eq.\eqref{z}, the full mass spectra of particles that make significant contributions to the effective potential.

\subsection{The effective potential and searching $S$}

The mass generation mechanism for particles in this model is still through the Higgs field like SM, however the Higgs field is always accompanied by the Dilaton field since the creation of non-zero VEV of the Higgs field requires a Dilaton field. This process is clearly shown in the minimization of the Higgs-Dilaton potential \cite{bars2000a,bars2001,survey,bars2006,kuo2006,33,pn}.

The effective one-loop potential is usually calculated in the manner of Jackiw \cite{cjd1,cjd2,cjd3} which is developed from the Coleman-Weinberg potential. The effective one-loop potential at 0K can be written as
\[
V^{0K}_\text{eff}(\phi_c) = \dfrac{\la_R}{4}\phi_c^4 - \dfrac{m_R^2}{2}\phi_c^2 + \La_R + \dfrac{1}{64\pi^2}
\sum\limits_{i = \mathcal{H},\mathcal{D},W,Z,t} n_i m_i^4(\phi_c)\ln \dfrac{m^2_i(\phi_c)}{v^2}\,,
\]
where $\la_R, m_R, \Lambda_R$ are renormalized parameters,  $m_i(\phi_c)$ is mass at tree level derived in previous subsection, i.e., SM particles and Dilaton;
$n_i$ with $i = \mathcal{D},\mathcal{H},W,Z,t$ are constants related to the degrees of freedom of each field and are given by
\[
n_{\mathcal{H}} = n_{\mathcal{D}} = 1, n_W = 6, n_Z = 3, n_t = -12 \, .
\]

To work out renormalized parameters, we have dropped the constant $\Lambda_R$, which results in the effective potential at 0 being 0, hence $V_{eff}(v)=V_{eff}(v)=0$. The familiar normalization conditions will be used,
\[
\begin{cases}
V_\text{eff}(v) = V_\text{eff}(0)=0 \, ,\\
V'_\text{eff}(v) = 0\, ,\\
V''_\text{eff}(v) = m_{\mathcal{H}}^2\,,
\end{cases}
\]
from which the coefficients in the effective potential are found out the following results
\[
\begin{cases}
\la_R = \dfrac{m_{\mathcal{H}}^2}{2v^2} - \dfrac{1}{16\pi^2 v^4}\sum\limits_{i} n_i m_i^4 \left(\ln\dfrac{m_i^2}{v^2}+\dfrac{3}{2}\right)\, , \\
m_R^2 = \dfrac{m_{\mathcal{H}}^2}{2} - \dfrac{1}{16\pi^2v^2}\sum\limits_i n_i m_i^4\, , \\
\La_R = \dfrac{m_{\mathcal{H}}^2v^2}{8} - \dfrac{1}{128\pi^2}\sum\limits_i n_i m_i^4\, .
\end{cases}
\]

\begin{figure}[h!]
	\centering
	\includegraphics[width=0.5\textwidth]{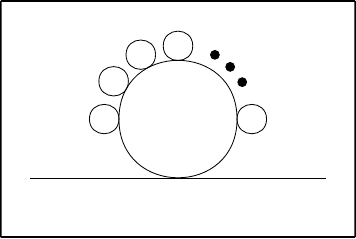}
	\caption{Contribution of $n$-loop daisy to the self-energy for the scalar field.}
	\label{fig18} 
\end{figure}

Another problem arises, the small value of $\phi_c/T_c$ causes the perturbation theory to break down due to infrared divergence at high temperatures and only non-perturbative calculations are acceptable. To solve this divergence problem we will add second order divergent daisy loops, that is, add n second order divergent loops as shown in Figure \ref{fig18}.

In case the daisy loops are taken into account, at the high temperature, the effective potential takes the form,
 \begin{equation}V_{\mathrm{eff}}(\phi_c,T)=\frac{\lambda(T)}{4}\phi_c^4+D(T^2-T_0^2)\phi_c^2-\frac{T}{12\pi}\sum_{\mathcal{H},W,Z,\mathcal{D}}g_i\left[\frac{m_i^2(v)\phi^2_c}{v^2}+\Pi_i(T)\right]^{\frac{3}{2}}.\label{11}\end{equation}
 If we do not consider daisy loops, Eq.\eqref{11} will not have $\Pi_i$ functions, the coefficients in Eq.\eqref{11} are of the form, 
 \begin{equation}\begin{gathered}
 		\lambda(T)=\frac{m_{\mathcal{H}}^{2}}{2v^{2}}+\frac{1}{16\pi^{2}v^{4}}\left(\sum_{i=\mathcal{H},\mathcal{D},W,Z}n_{i}m_{i}^{4}\ln\frac{A_{b}T^{2}}{m_{i}^{2}}+n_{t}m_{t}^{4}\ln\frac{A_{f}T^{2}}{m_{t}^{2}}\right), \\
 		D=\frac{m_{\mathcal{H}}^{2}+m_{\mathcal{D}}^{2}+6m_{W}^{2}+3m_{Z}^{2}+6m_{t}^{2}}{24v^{2}}, \\
 		T_{0}^{2}=\frac{1}{D}\left[ \frac{m_{\mathcal{H}}^{2}}{4}-\frac{1}{32\pi^{2}v^{2}}\sum_{i=\mathcal{H},\mathcal{D},W,Z,t}n_{i}m_{i}^{4} \right]. 
 \end{gathered}\end{equation}
 The contributions of daisy loops are stored in the following functions,
 \begin{align}
 	&\Pi_W(T)=\ \frac{22}{3}\frac{m_W^2}{v^2}T^2
 	,\label{dw}\\
 	&\Pi_Z(T)=\ \frac{22}{3}\frac{(m_Z^2-m_W^2)}{v^2}T^2
 	,\label{dz}\\
 	&\Pi_h(T)=\ \frac{m_{\mathcal{H}}^2+2m_W^2+m_Z^2+2m_t^2}{4v^2}T^2
 	,\label{dh}\\
 	&\Pi_d(T)\sim\frac{m_{\mathcal{D}}^2}{v^2}T^2.\label{dd}
 \end{align}
 
In Eq.\eqref{11}, $\ln A_b=3.907$ or $\ln A_f=1.135$, and there is an additional contribution of Dilaton, which will push the EWPT process to be more intense. As analyzed in the case without daisy loops in Ref.\cite{pn}, $300$ GeV $\le m_{\mathcal{D}} \le 650$ GeV, it will have the large enough EWPT strength ($S$). However, when the mass of Dilaton is too large, it will lead to large error in the effective potential. In Ref.\cite{pn} only $S$ is calculated but the convergence conditions and errors of the effective potential are not taken into account. The first we can see this, if the ratio between mass and temperature $(m/T)$ is not less than $2.2$ the potential difference will be in error by more than $5\%$ \cite{24}. The second, in Table \ref{vb}, when the Dilaton mass is larger than $550$ GeV, $\phi_c$ does not converge to $246$ GeV but exceeds $246$ GeV. 

$S=\frac{\phi_c}{T_c}$, is the ratio of VEV to the temperature at the time of phase transition. $T_c$ is called the critical temperature. With the equations from Eq.\eqref{11} to Eq.\eqref{dh}, we can probe $S$ for any given mass of Dilaton. The daisy loop contribution of Dilaton is ignored because its mass is larger than that of Top quark and at temperatures around $100-200$ GeV its contribution is small. Also the exact calculation of these contributions is quite complicated, we can only estimate them as Eq.\eqref{dd}. The equations from \eqref{dw} to \eqref{dh}, can be found in Refs.\cite{carrington,curtin,katz} for details, reduces the value of 3rd order component in Eq.\eqref{11}, leading to $S$ decreasing when there are daisy loops.

In Table \ref{vb}, $S$ with daisy loops is always smaller than one without daisy loops. However, this difference does not exceed $0.2$ or $10\%$. This also further confirms that the contribution of daisy loops in the critical temperature region is negligible but their contributions become significant in the higher one.

In summary, the first order EWPT has been recalculated in this section and then calculate the sphaleron energy and gravitational wave in the next section. The steps to find $S$ are as follows:

\begin{itemize}
	\item Choosing the Dilaton mass domain. Following Ref.\cite{pn}, we choose the Dilaton mass from $300$ GeV. Running the effective potential for different temperatures. $T_c$ is the temperature for which $v_{eff}(\phi_c, T_c)=0$.
	\item $\phi_c$ is the second non-zero VEV of $v_{eff}$. From there we calculate $S=\frac{\phi}{T_c}$. The convergence condition for $\phi_c$ is that it must be less than $246$ GeV.
\end{itemize}

\begin{table}[H]
	\centering
	\begin{tabular}{|c|c|c|c|c|c|c|c|c|c|}\hline\hline
		$m_{\mathcal{D}}$&$v_c$ &$T_c$&\multirow{2}{*}{$S$}&$E^T_{sph}$&$E_0$ &\multirow{2}{*}{$\alpha$}&\multirow{2}{*}{$\betah$}&\multirow{2}{*}{$\Omega h^2 (f_{peak})$}& \multirow{2}{*}{Daisy loops}\\
		
		[GEV]&[GeV]&[GeV]&&[GeV]&[GeV]&&&&\\
		\hline
		\hline
		\multirow{2}{*}{300}
		&	107.492	&	136.162	&	0.789439	&	7658.14	&	\multirow{2}{*}{9106.11}&	0.00766406	&	25.3093	&	6.34889$\times 10^{-16}$& without\\
		&	98.2028	&	146.281	&	0.671331	&	7636.04	&							&	0.00480408	&	23.4906	&	1.08379$\times 10^{-16}$&with\\
		\hline
		\multirow{2}{*}{350}
		&	133.417	&	129.726	&	1.02845		&	7751.48	&	\multirow{2}{*}{9094.9}&	0.0132065	&	26.8886	&	5.07648$\times 10^{-15}$& without\\
		&	128.617	&	138.449	&	0.928987	&	7727.12	&							&	0.00945377	&	25.1155	&	1.46259$\times 10^{-15}$& with\\
		\hline
		\multirow{2}{*}{400}
		&	160.455	&	122.702	&	1.30768	&	7905.02	&	\multirow{2}{*}{9079.14}&	0.0211503	&	28.991	&	2.958$\times 10^{-14}$& without\\
		&	158.371	&	130.329	&	1.21517	&	7888.4	&							&	0.0161664	&	27.2371	&	1.10564$\times 10^{-14}$& with\\
		\hline
		\multirow{2}{*}{450}
		&	190.502	&	115.103	&	1.65505	&	8113.7	&	\multirow{2}{*}{9055.43}&	0.0326829	&	31.7208	&	1.44943$\times 10^{-13}$& without\\
		&	189.376	&	121.835	&	1.55436	&	8106.3	&							&	0.0256766	&	29.9408	&	6.06993$\times 10^{-14}$& with\\
		\hline
		\multirow{2}{*}{500}
		&	229.877	&	108.027	&	2.12796	&	8392.06	&	\multirow{2}{*}{9020.63}&	0.0522199	&	34.9581	&	7.77137$\times 10^{-13}$& without\\
		&	227.515	&	113.854	&	1.99831	&	8384.29	&							&	0.0410284	&	33.1385	&	3.30201$\times 10^{-13}$& with\\
		\hline
		\multirow{2}{*}{550}
		&	287.102	&	106.276	&	2.70147	&	8792.76	&	\multirow{2}{*}{8970.03}&	0.0938256	&	37.2308	&	6.25938$\times 10^{-12}$& without\\
		&	281.168	&	110.832	&	2.53687	&	8769.76	&							&	0.0725315	&	35.6069	&	2.57741$\times 10^{-12}$& with \\
		\hline
		\multirow{2}{*}{600}
		&	348.246	&	114.415	&	3.04372	&	9264.72	&	\multirow{2}{*}{8895.83}&	0.158098	&	36.4387	&	3.90875$\times 10^{-11}$& without\\
		&	339.772	&	117.85	&	2.88309	&	9229	&							&	0.123424	&	35.2402	&	1.73786$\times 10^{-11}$& with \\
		\hline
		\hline
	\end{tabular}
	\caption{Results in the cases of Dilaton.}\label{vb}
\end{table}

 \section{Sphaleron and gravitational wave}\label{sec3}
 
 Under the conditions of Sakarov \cite{sakharov}, B violation must be mandatory and requires the existence of a sufficiently large sphaleron rate or sphaleron are relevant for the B violating processes. 
 
 The sphaleron energy functional consists of three components: the contribution of gauge fields, the Higgs kinetic energy and the effective potential.
 \be
 E_{sph}=\int dx^3 \left[ \fr{1}{4}W_{ij}^a W_{ij}^a +(D_i \phi)^\dagger (D_i \phi)+ V_{eff}\right].
 \label{e1}
 \ee
 
With the static field approximation (ie., $W^a_0=0$) and sphaleron in the spherically symmetric form \cite{11}:
 \be
 \left\{ \begin{array}{ll}
 	\phi(r)=\fr{v}{\sqrt{2}} h(r) i n_a \sigma^a \left(\begin{array}{cc} 0\\1\end{array} \right),\\
 	W_i ^a (r)=\fr{2}{g}\ep ^{aij} n_j \fr{f(r)}{r},
 \end{array}\right.
 \label{ham}
 \ee
 where $n_i \equiv \fr{x^i}{r}$ and $r$ is the radial coordinate in the spherical coordinates. $\sigma^a$ are the Pauli matrices and $a=1,2,3$. $\epsilon^{ij}$ is the Levi-Civita symbol.
 
 The non-zero temperature Sphaleron energy functional is written as follows
 \begin{equation}
 	E_{sph}^{T}=\fr{4 \pi v}{g} \int_0^\infty d\xi \left[ 4\left(\fr{df}{\xi}\right)^2
 	+\fr{8 f^2(1-f)^2}{\xi^2}  +\fr{\xi^2}{2}\left(\fr{dh}{\xi}\right)^2
 	+ h^2(1-f)^2 +\fr{\xi^2}{g^2 v^4}V_{eff}(h,T)\right],
 	\label{sphT}
 \end{equation}
 in which $g^2=\fr{G_F 8 m^2_W}{\sqrt{2}}$; $G_F=1.166 \times 10^{-15} \, \textrm{GeV}^{-2}$; $m_W=80.39\, \textrm{GeV}$, $\xi\equiv gvr$.
 
 Taking the variation of above function in terms of $h(\xi)$ and $f(\xi)$, we will obtain the equations of motion,
 \begin{align}
 	\xi^2\frac{d^2f}{d\xi^2}=2f(1-f)(1-2f)-\frac{\xi^2}{4}h^2(1-f),\label{21}\\
 	\frac{d}{d\xi}\left(\xi^2\frac{dh}{d\xi} \right)=2h(1-f)^2+\frac{\xi^2}{g^2v^4}\frac{\partial V_{eff}}{\partial h}. \label{22}
 \end{align}
 
 The contribution of effective potential to the sphaleron energy is less than $45\%$, but the effective potential acts as a source for bubbles (as seen from Eq.\eqref{21} and \eqref{22}).
 
\begin{figure}[H]
 	\centering
	\includegraphics[width=0.7\textwidth]{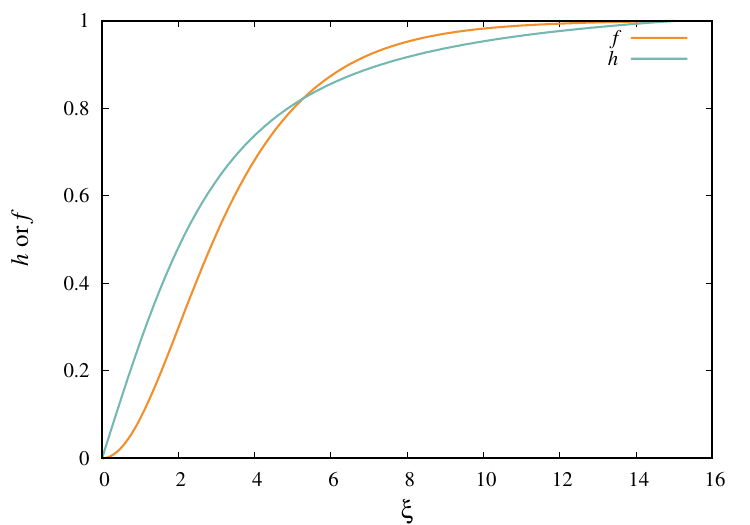} 
	\caption{Profile functions with $m_{\mathcal{D}}=300$ GeV.}\label{hinh1.1}
 \end{figure}

\begin{figure}[H]
 	\centering
 	\includegraphics[width=0.7\textwidth]{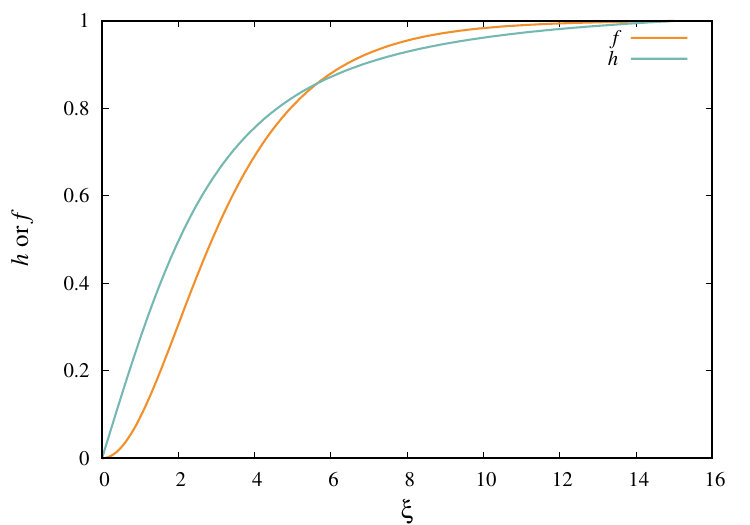} 
 	\caption{Profile functions with $m_{\mathcal{D}}=350$ GeV.}\label{hinh1.2}	
 \end{figure}

\begin{figure}[H]
 	\centering
 	\includegraphics[width=0.7\textwidth]{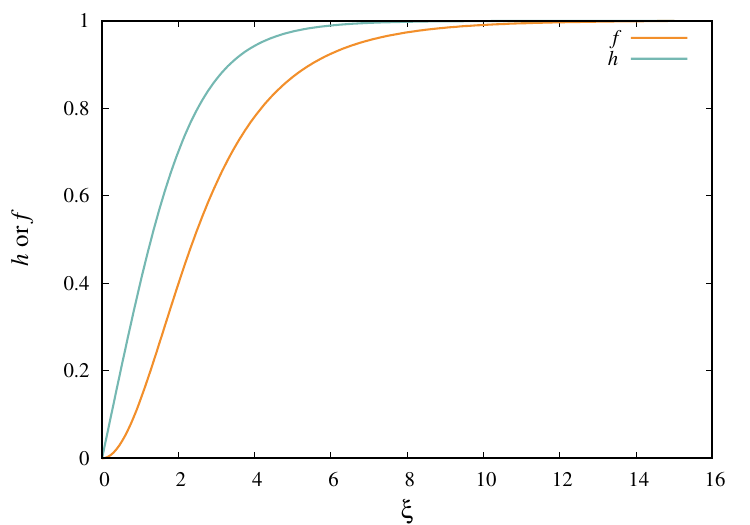}
 	\caption{Profile functions with $m_{\mathcal{D}}=550$ GeV.}\label{hinh1.3}
 \end{figure}

\begin{figure}[H]
 	\centering
 \includegraphics[width=0.7\textwidth]{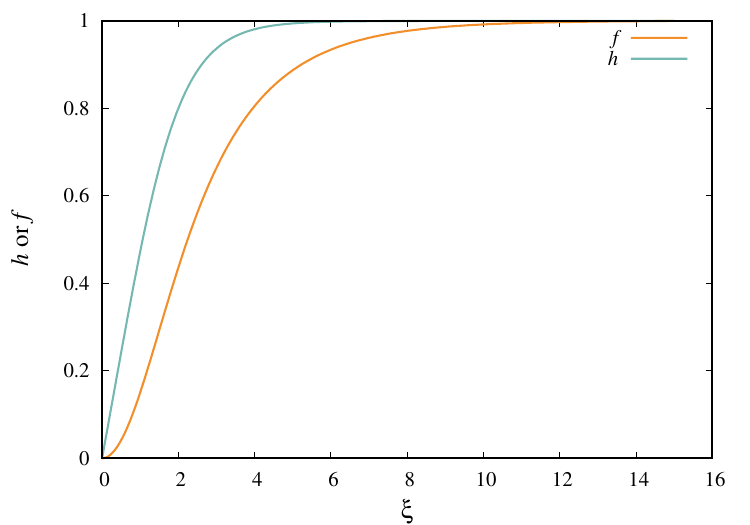}
 	\caption{Profile functions with $m_{\mathcal{D}}=600$ GeV.}\label{hinh1.4}
 \end{figure}

 Here, $h(\xi)$ and $f(\xi)$ have boundary conditions
 \be
 \left\{ \begin{array}{ll}
 	h(\xi\to 0)=f(\xi\to 0)=0,\\
 	h(\xi\to \infty)=f(\xi\to \infty)=1.
 \end{array}\right.
 \label{dk}
 \ee
These boundary conditions ensure that the sphaleron energy converges, because from Eqs. \eqref{22}, \eqref{21} and \eqref{sphT}, as $\xi$ increases, the functions $f(\xi),h(\xi)$ must approach 1 and the components in Eq.\eqref{sphT} must approach 0. For equations from \eqref{sphT} to \eqref{dk}, the numerical solutions are obtained as Figs. \ref{hinh1.1}, \ref{hinh1.2}, \ref{hinh1.3}, \ref{hinh1.4} and Table \ref{vb}.
 
 
 In Figs. \ref{hinh1.1}-\ref{hinh1.4}, the larger the mass of Dilaton is, the faster $f(\xi), h(\xi)$ approaches 1. This is consistent with the results in Table \ref{vb}, the larger the mass of Dilaton is, the larger $S$ is. When $m_{\mathcal{D}}<350$ GeV, $f(\xi)$ approaches 1 faster than $h(\xi)$ but when $m_{\mathcal{D}}>350$ GeV this approach to 1 happens in reverse. Because the larger $m_d$ is, the stronger the interaction between Higgs and Dilaton is, further delaying the expansion of Higgs bubble nucleation.
 
  \begin{figure}[H]
 	\centering
 	\includegraphics[width=0.7\textwidth]{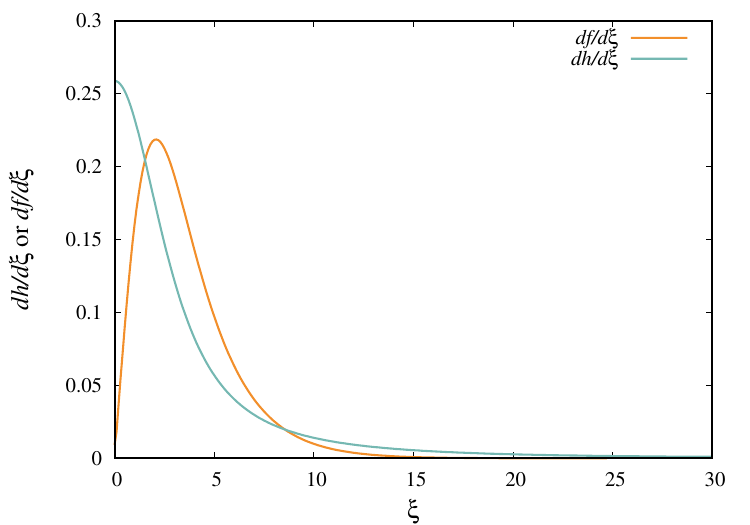}
 	\caption{$dh/d\xi$ and $df/d\xi$ with $m_{\mathcal{D}}=300$ GeV.}\label{hinh1.3boune}
 \end{figure}
 
 \begin{figure}[H]
 	\centering
 	\includegraphics[width=0.7\textwidth]{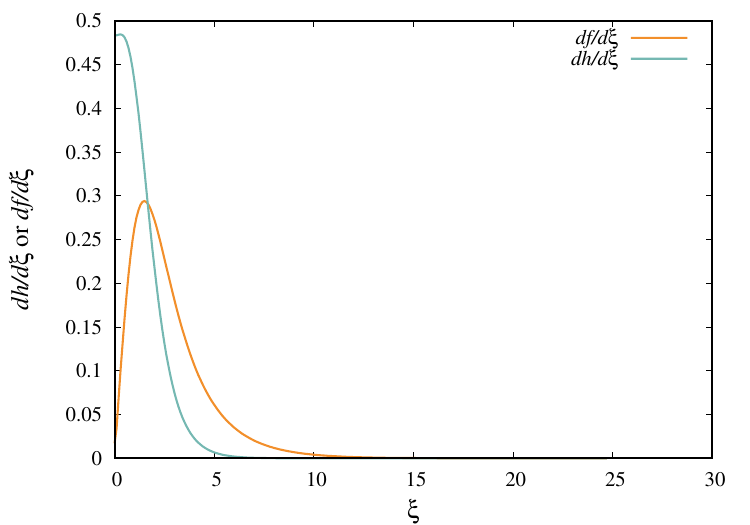}
 	\caption{$dh/d\xi$ and $df/d\xi$ with $m_{\mathcal{D}}=600$ GeV.}\label{hinh1.4bounce}
 \end{figure}
 
 Figs.\ref{hinh1.3boune} and \ref{hinh1.4bounce} show the expansion speed of bubbles (like the bounces of bubbles) corresponding to the solutions in Figs.\ref{hinh1.1} and \ref{hinh1.4}. $dh/d\xi$ always decreases as $\xi$ increases. $df/d\xi$ increases to a maximum value and then also decreases as $\xi$ increases. Additionally, this numerical calculation can also be done by referring to the packages in Ref.\cite{wain}.\\
 
 The largest sphaleron energy is about $9.2$ TeV with the Dialton mass of about $600$ GeV. But to avoid the VEV divergence problem, the Dilaton mass must be less than $550$ GeV, therefore the sphaleron energy in this model is about $8.4$ TeV. This energy range is about 1 TeV smaller than that in the SM.
 
 There are three processes which are involved in the production of GWs at a first-order PT: Collisions of bubble walls, sound waves in the plasma and Magnetohydrodynamic (MHD) turbulence \cite{71}. These three processes generically coexist, and the corresponding contributions to the stochastic GW background should linearly combine, at least approximately, so that \cite{71}	
 \be
 h^2\Om_{GW}=h^2\Om_{Coll}+h^2\Om_{sw}+h^2\Om_{tur}.\label{24}
 \ee	
 
The first parameter $\al $ is determined at the nucleation temperature as follows \cite{2gw, 2gwa,2gwb}:
\begin{align}
	\al &=\fr{\ep }{\rho_{rad}^*},\label{alls}\\
	\ep &=\left(V_{eff}(v(T),T)-T \fr{d}{d T}V_{eff}(v(T),T)\right)_{T=T_C},\label{alla}\\
	\rho_{rad}^*&=g^* \pi^2 \fr{T^4_c}{30}=106.75\pi^2 \fr{T^4_c}{30}.
	\label{all}
\end{align}

This parameter depends only on the configuration of effective potential. Thus in all models the parameter which is easily calculated by usual way when the effective potential is known, determines the strength of phase transition. 

There are three different scenarios depending on the bubble wall velocity \cite{71} which is defined via $\al $:
 \be
 v_b= \fr{\fr{1}{\sqrt{3}}+\sqrt{\al ^2+\fr{2\al }{3}}}{1+\al }.
 \label{al}
 \ee
 \begin{figure}[H]
 	\centering
 	\includegraphics[width=0.5\textwidth]{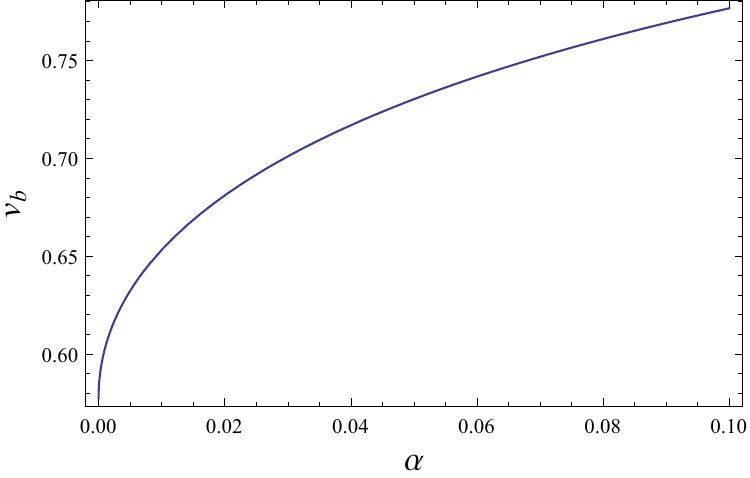}
 	\caption{The behavior of $v_b$ depends on $\alpha$. }\label{hvb}
 \end{figure}
 
 With the values of $\alpha$ in Table \ref{vb}, $0.009<\alpha <0.16$ and Fig.\ref{hvb} represents Eq. \ref{al}. Comparison between Table \ref{vb} and Fig.\ref{hvb}, we see that $v_b \approx 0.65-0.75$. It is not much smaller than 1 so that we approximate $v_b\sim 1$. If $v_b\ll 1$, we must have other approximations about $\Omega h^2$ \cite{71}. Therefore, the generation of gravitational waves has only two components: the sound wave and turbulence \cite{71, 73, 73a, 73b, 73c, 73d, 73e}. The gravitational wave energy density parameters of sound waves is defined as follows \cite{phu5gw, 71}:
 \begin{equation}
 	h^2\Omega_{sw}^2(f)=2.65\cdot10^{-6}v_b k_w^2
 	\frac{H^*}{\beta}\left( \frac{\alpha}{1+\alpha} \right)^2\left( \frac{100}{106.75} \right)^{1/3}S_{sw}(f),\label{26}
 \end{equation}
 where
 \begin{equation}
 	\begin{cases}
 		S_{sw}=\left( \frac{f}{f_{peak,sw}} \right)^3\left( \frac{7}{4+3\left( \frac{f}{f_{peak,sw}} \right)^2} \right)^{7/2}.\\
 		v_b\sim 1.\\
 		k_w=\alpha(0.73+0.083\sqrt{\alpha}+\alpha)^{-1}.
 	\end{cases}
 \end{equation}
 And GWs from turbulence in the cosmic fluid is given by
 \begin{equation}
 	h^2\Omega_{tur}^2(f)=3.35\cdot10^{-4}\left( \frac{k_t\alpha}{1+\alpha} \right)^{3/2}
 	\frac{H^*}{\beta}\left( \frac{100}{106.75} \right)^{1/3}S_{tur}(f),\label{28}
 \end{equation}
 in which 
 \begin{equation}
 	\begin{cases}
 		S_{tur}=\left( \frac{f}{f_{peak,tur}} \right)^3\left( \frac{1}{1+\left( \frac{f}{f_{peak,tur}} \right)} \right)^{11/3}\frac{1}{1+8\pi f/h^*}.\\
 		H^*=\frac{16.5\cdot 10^{-3} T_c}{100}\left( \frac{106.75}{100} \right)^{1/6}; v_b\sim 1.\\
 		k_t=0.05\alpha(0.73+0.083\sqrt{\alpha}+\alpha)^{-1.}\\
 		f_{peak,tur}=2.7\cdot10^{-2}\frac{\beta}{H^*}\frac{T_c}{100}\left( \frac{106.75}{100} \right)^{1/6}.
 	\end{cases}
 \end{equation}
 
$f_{peak, sw/tur}$ are the frequencies at which $\Omega h^2$ reache maxima. Because the effective potetial contributes about $45\%$ to the sphaleron energy. We have a possible approximation \cite{phong2022} for the second parameter (the timescale of phase transition),	
 \be
 \fr{\beta}{H^*}\approx \left[T \fr{d\left( S(t)\right)}{dT}\right]_{T=T_N}\approx\left[T \fr{d\left( S_3/T\right)}{dT}\right]_{T=T_N}\approx \left[T \fr{d\left(0.45\fr{E^T_{sph}}{T}\right)}{dT}\right]_{T=T_N}.
 \label{beta}
 \ee
 
$T_N$ in Eq. \eqref{beta} is a nucleation temperature which is usually smaller than $T_c$. Although at $T_c$, $S_3$ diverges \cite{moris}, the sphaleron energy can be determined. Therefore, we need to determine a nucleation temperature $T_N$ and then calculate $\fr{\beta}{H^*}$. Instead, we try to approximate $S_3$ with the sphaleron energy. To avoid supercooling, $T_N$ should not be too much larger than $T_c$ or the supercooling parameter $\delta_{sc,n}=\frac{T_c-T_N}{T_c}$ \cite{moris} must not be too big. Also because determining the nucleation temperature is quite cumbersome but it is negligibly smaller than $T_c$, so we use $T_c$ in the third approximation step in Eq.\eqref{beta}. 
	
The third approximation in Eq.\eqref{beta} is due to the fact that $S_3$ does not contain components of the gauge fields. In Table \ref{Et3}, the Higgs contribution ($E_{Higgs}^T$ is the component that contains only the $h$ function in Eq.\eqref{sphT}) to the total sphaleron energy is estimated to be about 45\%. 

The third approximation in Eq.\eqref{beta} is an imperfect approximation. It is just an estimate for determining $\fr{\beta}{H^*}$. Usually the value of this quantity is fixed at around $10-100$ \cite{71,moris} for calculating gravitational waves. Therefore the estimates in the third approximation also give acceptable values (specifically from 25 to 37, as in Table \ref{vb}). Also when we have calculated the sphaleron energy, Eq.\eqref{beta} is a convenient approximation or we are just trying to change the context from determining the pair $(T_N, S_3)$ to the pair $(T_c, E)$. This is a mathematical technique but it requires physical care.
\begin{table}[H]
	\centering
	\resizebox{0.9\textwidth}{!}{\begin{tabular}{|c|c|c|c|c|c|c|}
			\hline \hline
			\multirow{2}{*}{$m_{\mathcal{D}}[GeV]$}
			&\multicolumn{3}{c|}{With daisy loops}&\multicolumn{3}{c|}{Without daisy loops}\\
			\cline{2-7}
			&{$E^{T_c}_{Higgs}[GeV]$}&{$E^{T_c}_{sph}[GeV]$}&{$E^{T_c}_{Higgs}/E^{T_c}_{sph} (\%)$}&{$E^{T_c}_{Higgs}[GeV]$}&{$E^{T_c}_{sph}[GeV]$}&{$E^{T_c}_{Higgs}/E^{T_c}_{sph} (\%)$}\\\hline
			400&3691.73&7889.04&46.7957&3695.42&7905.67&46.7439\\\hline
			450&3740.08&8107.09&46.1334&3741.64&8114.49&46.1106\\\hline
			500&3805.29&8385.24&45.3809&3807.15&8393.02&45.3609\\\hline
			550&3910.02&8770.95&44.5793&3916.76&8793.96&44.5392\\\hline
			600&4058.26&9230.51&43.9657&4070.84&9266.26&43.9319\\
			\hline
			650&4213.12&9640.4&43.7027&4228.75&9678.84&43.6907\\\hline
			700&4354.95&9971.64&43.6734&4371.84&10008.8&43.6798\\			
			\hline
			\hline
	\end{tabular}}
	\caption{The ratio between the Higgs component and the total sphaleron energy.}\label{Et3}
\end{table}

\begin{figure}[H]
 	\centering \includegraphics[width=0.7\textwidth]{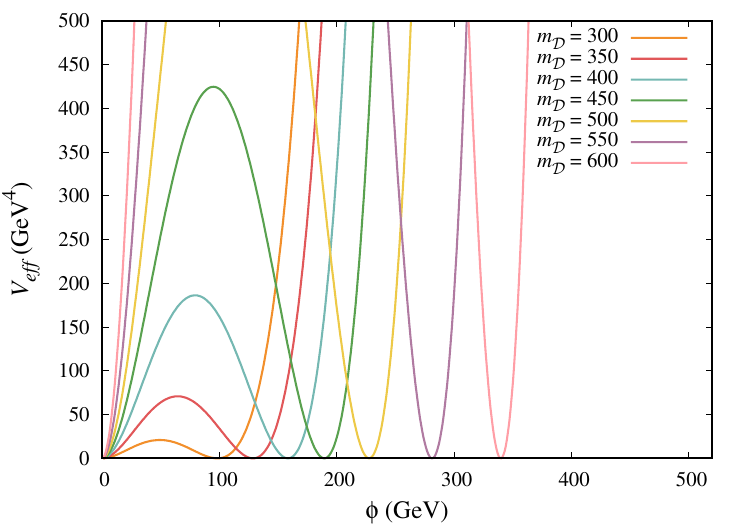}
 	\caption{The effective potential with $m_{\mathcal{D}}=300-600$ GeV.} \label{fig:2a}	
 \end{figure}

\begin{figure}[H]
 	\centering
 	\includegraphics[width=0.7\textwidth]{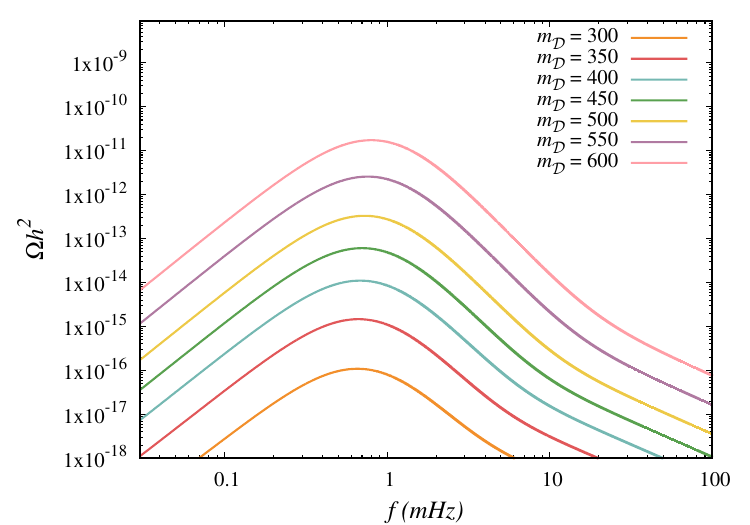} 
 	\caption{Gravitational wave energy density with $m_{\mathcal{D}}=300-600$ GeV.}\label{fig:2}
 \end{figure}

The contribution of daisy loops to the effective potential is negligible. Therefore, in this section, we will use the effective potential without the daisy loops to calculate the sphaleron and gravitational wave energies.  From equations \eqref{26}, \eqref{28} and related formulas, we solve numerically and obtain the results at $T_N\approx T_c$ in Table \ref{vb} and Figs. \ref{fig:2a}, \ref{fig:2}. From Table \ref{vb}, the sphaleron energies with and without daisy loop do not differ much.
 
 When the mass of Dilaton changes, the effective potential also changes so the sphaleron energy will change leading to $\Omega h^2$ depending on $m_{\mathcal{D}}$. In Fig.\ref{fig:2}, the larger $m_{\mathcal{D}}$ is, the higher the barrier of effective potential is, the higher the maximum of $\Omega h^2$ is. Because the higher the barrier is, the more violent the EWPT process is, so the gravitational waves generated from it are larger. In Fig. \ref{fig:2}, for $300$ GeV $\le m_{\mathcal{D}}<600$ GeV, the maximum value of $\Omega h^2$ lies in the range from $10^{-16}$ to $10^{-11}$. This is a very wide range, or in other words $\Omega h^2$ increases very rapidly over a mass range of $300$ GeV. Furthermore, for $\Omega h^2$ to increase 10 times, the mass of Dilaton increases by $50$ GeV.
 
 \begin{figure}[H]
 	\centering
  	\includegraphics[width=0.7\textwidth]{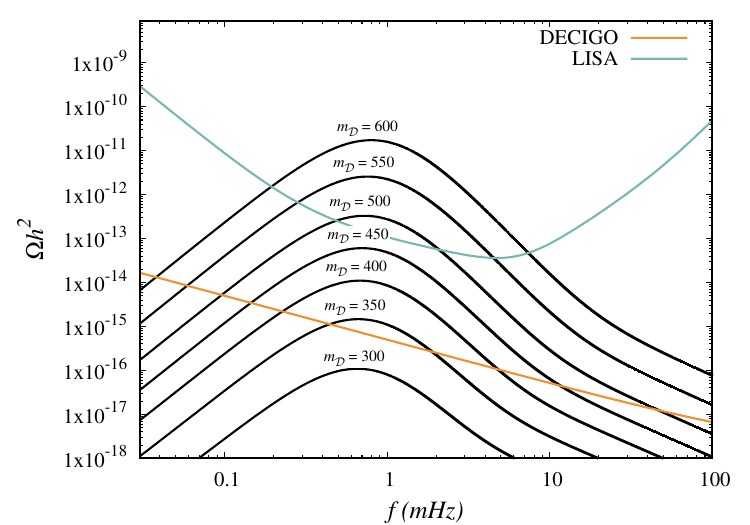}
 	\caption{With out the daisy loops, $m_{\mathcal{D}}=300-600$ GeV, gravitational wave energy density compared with the LISA and DECIGO data.}\label{dl}
 \end{figure}
 
  \begin{table}[!ht]
 	\centering
 	\caption{Sensitivity of proposed GW}
 	\begin{tabular}{|c|c|c|c|}\hline\hline
 		$f [mHz]$&$\Omega h^2$& Observatory & Ref.\\ \hline
 		\multirow{3}{*}{$0.02-0.12$}	
 		& $[0.02-1]\times 10^{-10}$ & LISA& Refs.\cite{phu1gw,kudoh,thranel}\\
 		& $[0.5-8]\times 10^{-14}$ & DECIGO & Refs.\cite{phu1gw,kudoh,thranel}\\
 		& --- & BBO & Refs.\cite{phu1gw,kudoh,thranel}\\
 		\hline\hline
 	\end{tabular}\label{dulieu}
 \end{table}	
 
 The region from $10^{-14}$ to $10^{-11}$ is quite important, corresponding to $400$ GeV $\le m_{\mathcal{D}}< 550$ GeV, because it can be included in the residual sensitivities of the experiment. Specifically, based on sensitivity data of future detector, in Table \ref{dulieu}, LISA and DECIGO can pick up the signal of gravitational waves generated by the EWPT process in this model as shown in Fig. \ref{dl}. Furthermore, as long as $m_{\mathcal{D}}<500$ GeV, gravitational waves from EWPT can be detected. The frequency domain for the maximum gravitational wave energy densities are $0.4-1.2$ mHz which lies within the detection range of LISA, DECIDO and BBO.\\
 
 In Table \ref{vb}, we have calculated that $\frac{\beta}{H*}=\left[ T \frac{d (0.45 \frac{E^T_{sph}}{ T})}{dT} \right]_{T=T_N}$ has values from 25 to 35 at temperatures from $120$ to $140$ GeV and $\gamma=\left[ T \frac{d (S_3/T)}{d T}\right] _{T=T_N}$ has values from 27 to 40. We find the difference between $\gamma$ and $\frac{\beta}{H*}$ to be in the range of $10\%$ to $20\%$. But the difference value still gives a value of $\Omega h^2$ within the detection range of LISA and DECIGO.

\section{Conclusion and discussion}\label{sec4}

The electroweak phase transition is triggered by the Dilaton to be of first order. The sphaleron energy and the gravitational wave energy density are calculated and shown to be compatible with future experimental sensitivities. Furthermore, the results in the article also indicate that the contributions of the daisy loops to the sphaleron energy are negligible.

\begin{table}[H]
	\centering
	\begin{tabular}{|c|c|c|c|c|c|c|c|c|c|c|}\hline\hline
		$m_{\mathcal{D}}$&$v_c$&$T_c$&$S$&$E^T_{sph}$&$E_0$ &\multirow{2}{*}{$E_0/v_0$}&\multirow{2}{*}{$E_c/v_c$}&\multirow{2}{*}{$\frac{|E_0/v_0-E_c/v_c|}{E_0/v_0}$}&\multirow{2}{*}{Decoupling}&\multirow{2}{*}{Daisy loops}\\
		
		[GeV]&[GeV]&[GeV]& &[GeV]&[GeV]& & & & &\\
		\hline
		\hline
		\multirow{2}{*}{300}
		&	107.492	&	136.162	&	0.789	&	7658.14	&	\multirow{2}{*}{9106.11}&\multirow{2}{*}{37.01}	&71.24	&92.48\%&57.09& without\\
		&	98.2028	&	146.281	&	0.671	&	7636.04	&	&	&77.75	&110\%&53.57&with\\
		\hline
		\multirow{2}{*}{350}
		&	133.417	&	129.726	&	1.028		&	7751.48	&	\multirow{2}{*}{9094.9}&\multirow{2}{*}{36.98}	&58.1	&57.11\%&59.78& without\\
		&	128.617	&	138.449	&	0.928	&	7727.12	&	&	&60.07	&62.43\%&56.17& with\\
		\hline
		\multirow{2}{*}{400}
		&	160.455	&	122.702	&	1.307	&	7905.02	&	\multirow{2}{*}{9079.14}&\multirow{2}{*}{36.91}	&	49.26&33.45\%&63.69& without\\
		&	158.371	&	130.329	&	1.215	&	7888.4	&	&	&49.8	&34.92\%&60.04& with\\
		\hline
		\multirow{2}{*}{450}
		&	190.502	&	115.103	&	1.655	&	8113.7	&	\multirow{2}{*}{9055.43}&\multirow{2}{*}{36.81}&42.59	&15.7\%&69.02& without\\
		&	189.376	&	121.835	&	1.554	&	8106.3	&	& &42.8	&16.27\%&65.27& with\\
		\hline
		\multirow{2}{*}{500}
		&	229.877	&	108.027	&	2.127	&	8392.06	&	\multirow{2}{*}{9020.63}&\multirow{2}{*}{36.67}		&36.5	&0.4\%&75.42& without\\
		&	227.515	&	113.854	&	1.998	&	8384.29	&	&&36.85	&0.49\%&71.59& with\\
		\hline
		\multirow{2}{*}{550}
		&	287.102	&	106.276	&	2.701	&	8792.76	&	\multirow{2}{*}{8970.03}&\multirow{2}{*}{36.46}		&30.61&16\%&79.74& without\\
		&	281.168	&	110.832	&	2.536	&	8769.76	&	&		&31.19	&14.45\%&76.34& with \\
		\hline
		\multirow{2}{*}{600}
		&	348.246	&	114.415	&	3.043	&	9264.72	&	\multirow{2}{*}{8895.83}&\multirow{2}{*}{36.16}		&26.6	&26.43\%&77.64& without\\
		&	339.772	&	117.85	&	2.883	&	9229	&	&	&27.16		&24.89\%&75.16& with\\
		\hline
		\hline
	\end{tabular}
	\caption{Results with the decoupling condition and the scaling law.}
\end{table}\label{vbsc}

The sphaleron scaling law \cite{scaling1,scaling2,scaling3}, is assumed to calculate the sphaleron energy at high temperatures when it is known at 0K ($E_0$ or $E(0K)$). It assumes that $E(0K)=\frac{v}{v_T}E(T)$. However, as the data in Table \ref{vbsc}, this law is not exact. Because $v_{eff}(0k)$ and $v_{eff}(T)$ are discontinuous when $T\longrightarrow 0$. $v_{eff}(T)$ is only valid for $T\neq 0$. This law can only be true in the temperature range from $T_N$ (the nucleation one) to $T_c$. Furthermore, from Table \ref{vbsc}, $E_0$ decreases very slowly, $E(T)$ increases rapidly corresponding to $300$ GeV $\le m_{\mathcal{D}} \le 600$ GeV. Because $f(\xi), h(\xi)$ at 0K approach 1 more slowly than they do at a non-zero temperature. However, this law is accurate for $m_{\mathcal{D}}=400$ GeV which may be the best value for Dilaton.

The decoupling condition \cite{decoupling, decouplingb, decouplingc}, the important one, which is related to the sphaleron rate and the expansion rate of Universe,

\begin{equation}\label{decoupling_equation}
	\frac{E_{sph}(T)}{T}-7ln\left(\frac{v(T)}{T}\right)+ln\left(\frac{T}{100 GeV}\right)>43.
\end{equation}

In Table \ref{vbsc}, the sphaleron energy at $T_c$ fully satisfies the condition in Eq.\eqref{decoupling_equation}. Although the right-hand side of Eq.\eqref{decoupling_equation} may be slightly changed, the values in the 10th column of Table \ref{vbsc} are considerably larger than $43$.

In the future, LISA, DECIDO and BBO can detect gravitational waves in the expected range. The GW of EWPT calculations could be a hint to indicate that there may be a contribution from the EWPT process in the detected wave spectrum. Additionally, the gravitational wave observations relevant to the early universe are summarized in Ref.\cite{gwhite}. Thus we have completely investigated the EWPT problem, calculating the Sphaleron energy and gravitational waves. 

The calculation of $\fr{\beta}{H^*}$ is only an estimate. If the mass of Dilaton is determined from other data channels, we will calculate $T_N$ and $S_3$ accurately and re-determine this parameter. This is an interesting future work for us.

In the 2T model, Dilaton is initially introduced to construct the 2T standard model through the $SP(2,R)$ symmetry, then we continue to analyze the role of Dialton in the EWPT problem. The standard model is added with a scalar singlet, it will be very difficult for us to explain or explain why we need to add a singlet (but there are many models built from the standard model with a scalar singlet). Thus, in terms of the dynamics of the EWPT problem, Dilaton in the 2T model is similar to a scalar singlet that is added in the standard model. However, in a larger context (i.e., in conjunction with Cosmology), Dilaton in the 2T model is directly connected to the 6-dimensional spacetime. In other words, the EWPT problem is explained by Dilaton in the 2T model, leaving us open that extra dimensions may exist. This is something that a pure scalar singlet fails to address.

\section*{ACKNOWLEDGMENTS}
This research is funded by University of Science, 
VNU-HCM under grant number T2024-10.

\end{document}